\documentstyle[12pt,epsf,axodraw]{article}

\def\bc{\begin{center}}
\def\ec{\end{center}}
\def\be{\begin{equation}}
\def\ee{\end{equation}}
\def\bea{\begin{eqnarray}}
\def\eea{\end{eqnarray}}
\def\ra{\rightarrow}

\def\kmung{$K\rightarrow\mu\nu\gamma$}

\def\G{G_{\mbox{\tiny F}} e V_{us}^\ast}
\def\red{$ R\!E\!D\!U\!C\!E$}
\def\eps{\varepsilon^{\mu\nu\alpha\beta}}

\def\cM{{\cal M}}
       
\begin{document}

\bc
{\huge \bf The Uses of Covariant Formalism \\
\mbox{for Analytical Computation of Feynman} \\[2mm]
 Diagrams with Massive Fermions}\\[5mm]

{\large R.N. Rogalyov\footnote{E-mail: rogalyov@mx.ihep.su},\\
Institute for High Energy Physics, Protvino, Russia.\\[8mm]
}

{\bf Abstract} 
\ec

{\noindent \small The bilinear combination of Dirac spinors 
$u(p_1,n_1)\bar u(p_2,n_2)$ is expressed in terms
of Lorentz vectors in an explicit covariant form.
The fact that the obtained expression involves only one auxiliary vector 
makes it very convenient for analytical computations 
with REDUCE (or FORM) package in the helicity formalism. 
The other advantage of the proposed formulas 
is that they apply to massive fermions as well as to massless fermions. 
The proposed approach is employed for the computation of one-loop 
Feynman diagrams and it is demonstrated that it considerably reduces 
the time of computations. 
}

\vspace*{7mm}

\section{Introduction}

Over many years, expressions of the type
\be\label{eq:BilinCombInitial}
w(p_1,n_1)\bar w(p_2,n_2) = S + V_\mu\gamma^\mu +T_{\mu\nu}\sigma^{\mu\nu}
 + A_\mu\gamma^\mu\gamma^5 + P\gamma^5
\ee
have received considerable study \cite{Bellomo}--\cite{Rogalyov}. They have been 
extensively used both in calculations of multiparticle
helicity amplitudes and in the analysis of polarization phenomena.
The method of computation of multiparticle helicity amplitudes 
(hereafter referred to as the spinor formalism) is based on
the representation of the polarization vectors of gauge bosons
in terms of spinors. It has been successfully applied to 
the computations of tree amplitudes of the reactions involving massless 
particles \cite{Gausmaecker} and has also been extended to the massive case
\cite{Ballestrero, GPS}.
However, a computation of a loop diagram with this method presents a challenge.

Another approach is to express the relevant products of spinors
in terms of Lorentz vectors (the method of calculations 
with the use of such expressions is referred to as the helicity formalism). 
However, the expressions
for $S, V_\mu$~{\it etc.} obtained in \cite{Fearing, Kriv}
are very cumbersome and 
involve singularities that hinder employing these expressions
in computations. One can use a particular basis to
avoid the singularities; in this case, there
emerges an additional auxiliary vector \cite{Galynskii, Rogalyov}. 

In this work, we discuss properties of these auxiliary vectors and
demonstrate that an expression of these vectors in terms of momenta
simplifies the computations and makes it possible to employ
the helicity formalism for an analitycal computation of the 
helicity amplitudes in the one-loop approximation.

\section{Basic Formulas}

In order to compute a complete set of the helicity amplitudes 
of some process involving 2 fermions, it is sufficient to
consider the case when the momenta $p_1$ and $p_2$ and the spin vectors 
$\pm N_1$ and $\pm N_2$ of these fermoins lie in the one 2-plane; in this case, 
polarization vectors of the fermions can be expressed in terms of the momenta
by the formulas
\be\label{eq:App2-DefPolVect}
N_1 = \frac{1}{\Delta}\left(\frac{p_1\cdot p_2}{m_1}p_1-m_1p_2 \right),\ \ \ \ \ 
N_2 = \frac{1}{\Delta}\left(-m_2p_1+\frac{p_1\cdot p_2}{m_2}p_2 \right),
\ee
where $m_1^2=p_1^2,\ m_2^2=p_2^2$, and $ \Delta= \sqrt{(p_1\cdot p_2)^2-m_1^2m_2^2}$. 
This assertion stems from
the fact that an arbitrary polarization state of a fermion is
a quantum-mechanical superposition of the states $|N\rangle$
and $|-N\rangle$, where $|N\rangle$ is the state with the spin
vector directed along some vector $N$ and $|-N\rangle$ is the
state with the spin vector directed oppositely to $N$.
For this reason, we restrict our attention to the basis
composed of the fermion states $|p_1, N_1 \rangle$,
$|p_1,-N_1 \rangle$, $|p_2, N_2 \rangle$, and $|p_2,-N_2 \rangle$
(or, in another notation, $u(p_1, N_1)$, $u(p_1,-N_1)$, $u(p_2, N_2)$, $u(p_2,-N_2)$),
where the vectors $N_1$ and $N_2$ are defined by (\ref{eq:App2-DefPolVect});
this basis is named `diagonal spin basis' \cite{Galynskii, Rogalyov}.
\bc
\begin{picture}(500,305)(0,0)
\LinAxis(50,20)(450,20)(2,2,2,0,1)
\LinAxis(250,20)(250,305)(2,2,2,0,1)
\Line(250,20)(50,220)
\Line(250,20)(450,220)
	\LongArrow(250,20)(350,148)
	\LongArrow(250,20)(378,120)
		\LongArrow(250,20)(150,244)
		\LongArrow(250,20)(138,70) 
	\LongArrow(250,20)(100,170)
	\LongArrow(250,20)(400,170) 
\Curve{(50,235)(150,148)(200,114)(250,100)(300,114)(350,148)(450,235)}
\Curve{(50,303)(150,244)(200,226)(250,220)(300,226)(350,244)(450,303)}
	\Text(112,78)[lb]{\Large $N_2$}
	\Text(145,268)[lb]{\Large $p_2$}
	\Text(335,162)[lb]{\Large $p_1$}
	\Text(375,98)[lb]{\Large $N_1$}
	\Text(413,233)[lb]{\Large $m_1$}
	\Text(410,300)[lb]{\Large $m_2$}
\Text(80,162)[lb]{\large $k_2$}
\Text(410,162)[lb]{\large $k_1$}
  \Text(256,295)[lb]{\Large \bf E}
  \Text(435,29)[lb]{\Large \bf p}
\end{picture}
\ec
\nopagebreak
\bc
Figure 1: \sl Momenta $p_1, p_2$ and vectors of spin $N_1, N_2$ lie in the one 2-plane
\ec

In this basis,
the combination of Dirac spinors  $u(p_1,N_1)\bar u(p_2,N_2)$,
where $p_1$ and $p_2$ are the momenta 
and $N_1$ and $N_2$ are
the polarization vectors, has the form \cite{CORE}
\bea\label{eq:App2-Main1}
u\bar u(\pm,\pm)=\left(j_1\frac{1\pm \gamma^5}{2} - j_2 \frac{1\mp\gamma^5}{2}+m_1\hat k_2-m_2\hat k_1\right)
\frac{\hat \omega_{\pm}}{\sqrt{2}}, \\  
\label{eq:App2-Main2}
u\bar u(\pm,\mp) = (j_1\hat k_1+m_1)\hat k_2\frac{1\pm \gamma^5}{2} + 
 (j_2\hat k_2+m_2)\hat k_1\frac{1\mp \gamma^5}{2}, 
\eea 
 where $u\bar u(+,-)\equiv u(p_1,n_1\!\!=\!\!+N_1) \bar u(p_2,n_2\!\!=\!\!-N_2)$ {\it etc}.,
\be\label{eq:App2-j1j2}
j_1=\frac{\sqrt{p_1\cdot p_2+m_1m_2}+\sqrt{p_1\cdot p_2-m_1m_2}}{\sqrt{2}},\ \ \ \ 
j_2=\frac{\sqrt{p_1\cdot p_2+m_1m_2}-\sqrt{p_1\cdot p_2-m_1m_2}}{\sqrt{2}},
\ee
\noindent and the light-like vectors $k_1, k_2$, and $\omega_\pm$ are given by
\be\label{eq:App2-k1k2}
k_1=\frac{1}{2\Delta}\left(j_1p_1-\frac{m_1}{m_2}j_2p_2\right),\ \ \ 
k_2=\frac{1}{2\Delta}\left(-\frac{m_2}{m_1}j_2p_1+j_1p_2\right),\ \ \ 
k_1\cdot k_2 = \frac{1}{2},
\ee
\begin{equation}\label{eq:ExplicitOmega2}
\hat \omega_\mp = -\frac{1}{\sqrt{2\Delta_3}}
\left(\hat k_1 \hat q \hat k_2 \frac{1\pm \gamma^5}{2} + \right.
\left. \hat k_2 \hat q \hat k_1 \frac{1\mp \gamma^5}{2} \right),
\end{equation}
where
\be\label{eq:Delta3}
\Delta_3=\delta^{k_1k_2q}_{k_1k_2q} = k_1\cdot k_2( 2\> k_1\cdot q\ k_2\cdot q - q\cdot q\ 
k_1\cdot k_2) 
\ee
and $q$ is an arbitrary vector such that $\Delta_3\neq 0$\footnote{Here
and below, $\gamma^5 =  i\gamma^0 \gamma^1 \gamma^2 \gamma^3$ and
$\varepsilon^{0123}=+1$. In this case, $T\!r \gamma^\mu \gamma^\nu 
\gamma^\alpha \gamma^\beta \gamma^5 = -4i\eps $ and
the operator $g(l,a)$ in the \red package (symbol $l$ marks the fermion line) 
should be identified with $ -\gamma^5$.}. To simplify computations, 
one should choose vector $q$ such that either $q\cdot q =0$ or $q\cdot k_1=q\cdot k_2 =0$.
In spite of the
presence of an arbitrary vector $q$ in (\ref{eq:ExplicitOmega2}), 
$\omega_+$ and $\omega_-$ are almost independent of $q$: they only acquire
a phase as the vector $q$ varies. 

The explicit expressions (\ref{eq:ExplicitOmega2}) for $\omega_\pm$
allow to compute the complete set of the helicity amplitudes 
using only momenta of the particles, that is, without resort to 
specific polarization vectors \cite{Gausmaecker} 
or `fundamental' spinors \cite{Ballestrero}. Such computation 
is possible due to the arbitrariness in a choice of the `gauge'
vector $q$. To explain this in more detail, let us consider a
diagram involving a fermion line of the type ${\cal F} =\bar u(p_2,n_2)
{\cal O} u(p_1,n_1)$, where ${\cal O}$ is something like $T$ in
formula (\ref{eq:InteKmung}) below. Since ${\cal F} =T\!r\, 
{\cal O} u(p_1,n_1)\bar u(p_2,n_2)$, it can be evaluated with the formulas
(\ref{eq:App2-Main1})--(\ref{eq:ExplicitOmega2}); this being so,
the momentum of {\bf any} particle other than a fermion of
momentum $p_1$ or $p_2$ can be inserted in (\ref{eq:ExplicitOmega2})
for vector $q$.

{\small The only exception from this rule is the decay $f_1\ra f_2 \gamma$,
where $f_1$ and $f_2$ are heavy and light fermions and $\gamma$ is the photon
(or gauge boson). In this case, there is no momentum independent of the fermion
momenta and, therefore, vector $q$ with the required properties does not exist. 
In the analysis of such reactions, the vectors $\omega_\pm$ should be
identified with the polarization vectors of the photon.}

\section{An Example of Computations}

Our approach is based on the representation of spinors in terms of vectors,
its application to loop computations is straightforward. Therein lies 
the difference between the proposed approach and that by
Ballestrero {\it et al.} \cite{Ballestrero}, in which all relevant
vectors should be expressed in terms of spinors. We illustrate 
the proposed method by computing the imaginary part of the one-loop diagram for the 
decay $K(p_K) \ra \mu(k)\nu(k')\gamma(q)$  (see Fig.~2). 
The kinematical variables here are defined as follows:
\bea\label{eq:VarPeng}
&& x=\frac{2p_K\cdot q}{M_K^2};\ \ y =\frac{2p_K\cdot k}{M_K^2};\ \ \ 
 \rho=\frac{m_\ell^2}{M_K^2}; \ \ \\ \nonumber
&&\lambda=\frac{1-y+\rho}{x};\ \   \ \zeta = (1-\lambda)(1-x)-\rho, \nonumber
\eea
where $m_\ell$ is the lepton mass and $M_K$ is the kaon mass.

We employ the Cutkosky rules \cite{Zuber} to replace cut propagators 
by the $\delta$ functions. Thus we obtain 
the expression for the imaginary part of the amplitude,
\be\label{eq:InteKmung}
\cM''={\alpha F \over 2\pi}\G \int dr { \delta(r^2-m_\ell^2) \delta \left((k+q-r)^2\right)
\over Z(r\cdot q, r\cdot k)} 
\, \bar u(k')(1+\gamma^5)T(r,k,k',q,\epsilon)v(k),
\ee
where $ T=8 (\hat k + \hat q)(1-\gamma^5) (m_\ell-\hat q - \hat k) \gamma^\nu
(m_\ell-\hat q- \hat r) \hat\epsilon (m_\ell-\hat r) \gamma^\nu$,
$Z$ is the product of the remaining (uncut) propagators, $r$ is the loop momentum, 
spinors $u$ and $v$ describe the neutrino and muon, and $\hat \epsilon = \epsilon_\mu (\pm)\gamma^\mu$.
Here
\be\label{EpsilonKmung}
\epsilon_\mu(\pm) = {\sqrt{2} \over 2M_K x \sqrt{\lambda\zeta}}
\left(-x\lambda k_\mu+x(1-\lambda)k'_\mu -(1-\rho-x)q_\mu \mp {2i\over M_K^2}\varepsilon_{kk'q\mu}\right).
\ee
is the polarization vector of the photon. The factor ${\alpha F \over 2\pi}\G$ 
in formula (\ref{eq:InteKmung}) should be thought of as an
effective coupling constant.

\setcounter{figure}{1}
\bc
\begin{figure}[thh] \hbox{
\hspace*{25pt}
       \epsfxsize=310pt \epsfbox{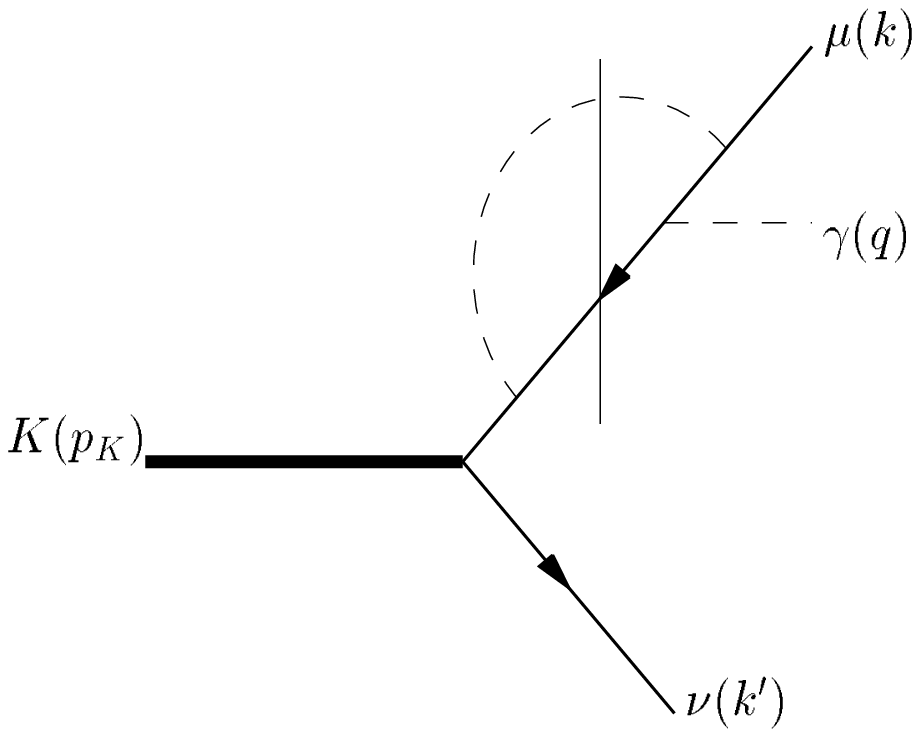} \hspace*{10pt}
       } \label{fig:2}
\caption{\sl Diagram giving a contribution to the 
amplitude of the decay \kmung.}
\end{figure}
\ec

Our method is based on computation of the helicity 
amplitudes.\footnote{Here there are four helicity amplitudes: 
the muon helicity (we select the reference frame
comoving with the center of mass of the muon and neutrino) takes one of the
two values $\pm 1/2$; photon helicity in any reference frame is $\pm 1$.}
Making use of the formulas
\bea\label{eq:BilinCombKmung}
&& \hspace*{-8mm} v_\mu(k,-N) \bar u_\nu(k') \! = \! {(\hat k - m_\ell) \hat k' \over
2 M_K \sqrt{1-x-\rho}} (1+\gamma^5), \\ \nonumber
&& \hspace*{-8mm} v_\mu(k,N) \bar u_\nu(k') \! = \! {M_K^2 (1-x-\rho) - m_\ell \hat k' \!\!\!
\over 2 M_K \sqrt{1-x-\rho}}\; \hat \omega_- (1+\gamma^5), 
\eea
where
\be\label{eq:OmegaKmung}
\hat \omega_- = - {\sqrt{2} \over 2M_K^3 x \sqrt{\lambda\zeta}}
\left(\hat k \hat q \hat k' (1+\gamma^5)+\hat k' \hat q \hat k (1-\gamma^5)-{2 M_K^2 \rho x \lambda \over 1-x-\rho} \hat k' \right)  
\ee
and
\be
N_\nu = {(1-x-\rho) k_\nu - 2\rho k'_\nu \over m_\ell (1-x-\rho)},
\ee
we express the quantity $\bar u(k')T(r,k,k',q,\epsilon)v(k)$ 
(which is nothing but the respective helicity amplitude) 
in the integrand in (\ref{eq:InteKmung}) in terms of the scalar products of 
$k,k',q$, and $r$ for each fixed polarization state of the muon and photon
{\bf and only then} perform integration with respect to $r$. 
The two integrations are trivial due to the $\delta$ functions,
the integtrand takes the form of a rational function $F=F(x, \lambda, 
\rho, v, \cos\phi)$:
\be\label{KmungIntdv}
\int dr \delta(r^2-m_\ell^2) \delta \left((k+q-r)^2\right) F(r) = 
{ x(1-\lambda) \over 8\tau} \int_{-1}^{1} dv \int_{0}^{2\pi} d\phi F,
\ee
where \ \
\be
v={4\tau \over x^2(1-\lambda)^2}\;{q\cdot r\over M_K^2}-
{2\rho \over x(1-\lambda)} -1,
\ee
and the azimuth angle $\phi$ specifies the direction 
of the projection of the vector $r$ onto the plane orthogonal 
to the 4-vectors $k$ and $q$. 
The function $F(x, \lambda, \rho, M_K/M_\pi ,v, \cos\phi)$ 
is a polynomial in $\cos\phi$, therefore,
integration with respect to
$\phi$ is trivial; this being so, the integrand in the integral with respect to $v$
has the form $A(v)+B/(Cv+D)$, where $A(v)$ is a polynomial in $v$ and
$B, C, D$ are independent of $v$. 
The computation of the diagram in Fig.~2 is made with the {\it REDUCE} package
\cite{REDUCE}.
This diagram is calculated exactly, no approximation is used \cite{Likhoded}.
A computation in the helicity formalism is by an order of magnitude faster than
a computation with the traditional method.

Now we present the contribution of the diagram in Fig.~2 
to the imaginary part of the helicity amplitudes
as it was calculated using the above procedure:

\begin{verbatim}
%%% The variables used below are defined as follows:
let w=1-lmbd;
let b^2=x*lmbd;
let c^2=x*(1-lmbd);
let az^2=1-rho-x;
let d^2=c^2*az^2-rho*b^2;

%%% Results of computations:
GammaMinusMuMinus := (16*rho**3*me*mk**4*w*x**2*( - w + 1) 
          + 16*rho**2*me*mk**4*w*x**2*( - w**2*x + w*x + w - 1) 
          + 4*rho*me*mk**4*w**2*x**4*(w**2 - 1) 
          + 8*me*mk**4*w**3*x**4*(w*x - w - x + 1))
/(rho*az*b*d + az*b*d*w*x)$

GammaMinusMuPlus := (16*rho**2*mk**5*w*x**2*(w - 1) 
		+ 4*rho*mk**5*w**2*x**3*(w - 1))
/(rho*az + az*w*x)$

GammaPlusMuMinus := (16*rho**3*me*mk**4*w*x**2*( - w + 1) 
           + 8*rho**2*me*mk**4*w*x**2*( - 2*w**2*x - w*x + 2*w + 3*x - 2) 
	   + 4*rho*me*mk**4*w**2*x**3*(w**2*x - 9*w*x + 6*w + 8*x - 6) 
	   + 4*me*mk**4*w**3*x**4*( - w*x + w + x - 1))
/(rho*az*b*d + az*b*d*w*x)
	+ ln(w*x/rho+1) * (8*rho*b*d*me*mk**4)/az$

GammaPlusMuPlus := (8*rho**2*mk**5*w*x**2*(w + 1) 
          + 4*rho*mk**5*w**2*x**2*( - w*x + 6*x - 4) 
	  + 4*mk**5*w**3*x**3*(x - 1))
/(rho*az + az*w*x)
	+ ln(w*x/rho+1) * ( - 8*rho**2*mk**5*x + 8*rho*mk**5*w*x*( - x + 1))/az$,
\end{verbatim}

\noindent where {\tt lmbd}~$=1-\lambda$, {\tt rho}~$=\rho$, {\tt me}~$=m_\ell$,
{\tt mk}~$=M_K$, the variable {\tt GammaPlusMuMinus}~is the imaginary part
of the amplitude for the photon helicity $+1$
amd the muon helicity $-1/2$ (in the reference frame comoving with the
center of mass of the lepton pair) {\it etc.} 
Here we avoid cumbersome denominators by expressing all vectors in terms of
the light-like vectors $q$,
\be 
k_1={1\over M_K\,\sqrt{1-x-\rho}}\;\left(k-\;{\rho\over 1-x-\rho} k' \right),\ \ 
\mbox{and} \ \ k_2={k'\over M_K\,\sqrt{1-x-\rho}} 
\ee
from the outset.\\[2mm]

{\large \bf Note added}\\[1mm]

\noindent The imaginary part of the above diagram is needed
for the determination of the transverse component of the muon
spin, which may provide an indicator of $C\!P$ violation.
It should be noted that, using the relations similar to (\ref{eq:App2-Main1})
and (\ref{eq:App2-Main2}),
the average value of the transverse component of the 
muon spin\footnote{It measures a half of the muon polarization} 
in the reaction under consideration can be expressed 
in terms of the helicity amplitudes as follows:
\bea
\label{TrSpinKmung}
\xi ={1 \over {\cal N}^2}
\left( \cM_{-\; -}' \cM_{-\; +}'' - \cM_{-\; +}' \cM_{-\; -}'' \right.
\left. + \cM_{+\; -}' \cM_{+\; +}'' - \cM_{+\; +}' \cM_{+\; -}'' \right),
\eea
where ${\cal N}$ is the normalization factor, ${\cal N}^2=\sum_{i,j=\pm}|\cM_{i,j}|^2$,
and $\cM_{r,s}'$ and $\cM_{r,s}''$ are the real and imaginary parts
of the helicity amplitudes, $\cM_{r,s} = \cM_{r,s}' + i\cM_{r,s}''$
(indices $(r,s=\pm)$ mark the photon and muon helicities).

\vspace*{5mm}
{\Large \bf Appendix. Properties of the Vectors $\omega^\pm_\mu$.}
\vspace*{3mm}

Any ordered pair of real light-like vectors $k_1$ and $k_2$: 
$k_1\cdot k_2=\frac{1}{2}$ determines the other pair of the complex zero-norm
vectors $\omega_+$ and $\omega_-$, which are complex-conjugated to each other
and satisfy the relations
\be\label{eq:1314}
\omega_+\cdot k_1=\omega_+\cdot k_2=0,\ \ \ \omega_-\cdot k_1=\omega_-\cdot k_2=0,\
\ \ \ \omega_-\cdot\omega_+=-1.
\ee
Now, let the tensors
\[ k_1^\mu \omega_+^\nu -\omega_+^\mu k_1^\nu \ \ \ \ \mbox{and}\ \ \ \  
k_2^\mu \omega_-^\nu - \omega_-^\mu k_2^\nu \]
be self-dual:
\begin{equation}\label{eq:17}
\hat k_1 \hat \omega_+(1-\gamma^5) = \hat k_2 \hat \omega_- (1-\gamma^5) = 0, 
\end{equation}
whereas the tensors 
\[k_2^\mu \omega_+^\nu -\omega_+^\mu k_2^\nu \ \ \ \ \mbox{and} \ \ \ \ 
k_1^\mu \omega_-^\nu -\omega_-^\mu k_1^\nu \] 
be anti-self-dual:
\begin{equation}\label{eq:18}
\hat k_2 \hat \omega_+ (1+\gamma^5) =\hat k_1 \hat \omega_- (1+\gamma^5) = 0. 
\end{equation}
{\bf The vectors $\omega_+$ and $\omega_-$ are determined by the pair
$k_1,\ k_2$ and the requirements (\ref{eq:1314}) and (\ref{eq:17})--(\ref{eq:18}) 
up to a phase.}

\begin{itemize}
\item{The vectors $\omega_+ \ (\omega_-)$ can be interpreted as the
polarization vectors of the photon of momentum $k_1$ and {\it positive
(negative)} helicity (in the gauge $k_2\cdot A=0$)}
\end{itemize}

The vectors $\omega_\pm$ can be represented in the explicit 
form\footnote{Note that $\gamma^5 =  i\gamma^0 \gamma^1 \gamma^2 \gamma^3$ and
$\varepsilon^{0123}=+1$.} \cite{Rogalyov, CORE}
\begin{equation}\label{eq:OmegaExplicit}
\omega^\mu_\pm \ =\ \frac{1}{\sqrt{2\Delta_3}}\> (k_1\cdot k_2 q^\mu -
q\cdot k_2 k_1^\mu -q\cdot k_1 k_2^\mu \pm i\varepsilon^{\mu q k_1 k_2}),
\end{equation}
where $\Delta_3$ is defined in (\ref {eq:Delta3}).

It should also be mentioned that, in the framework of the Cartan theory
of spinors \cite{Cartan}, a Dirac spinor in 4-dimensional space can be 
interpreted as an isotropic 2-plane in 5-dimensional pseudo-Euclidean space
(4 dimensions in it are associated with the Minkowski space-time
and the 5{\it th} dimension is associated with $\gamma^5$
giving the signature $(+,+,-,-,-)$). This being so,
such plane for the spinor\footnote{Solution of the Dirac equation with 4-momentum $p$
($p^2=m^2$) and vector of spin $n$ ($n^2=\,-\,1$).}
u(p,n) is nothing but the linear span of the zero-norm 5-vectors $(1,n)$ and
$(0,\omega_+)$, where the 4-vector $\omega_+$ is associated with
the vectors 
\[ k_1={1\over 2}\;\left({p\over m}+n\right) \ \ \mbox{and} \ \ \ 
k_2={1\over 2}\;\left({p\over m}-n\right) \]
according to the above procedure.


\begin{thebibliography}{99}
\bibitem{Bellomo} E. Bellomo, {\it Nuovo Cim.} {\bf 21}, 730 (1961);\\
A.A. Bogush, F.I. Fedorov, {\it Vesti AN BSSR} ü2 (1961), p.26.
\bibitem{Gausmaecker} P.de Gausmaecker {\it et al., Nucl.Phys.}
{\bf B206}, 53 (1982);\\
F. Berends, W. Giele, {\it Nucl.Phys.} {\bf B294}, 700 (1987);\\
F.Berends {\it et al., Nucl.Phys.} {\bf B357}, 32 (1991).
\bibitem{Fearing}  H.W. Fearing, R.R. Silbar, {\it Phys. Rev.} {\bf D66}, 471 (1972).
\bibitem{Kriv} M.I. Krivoruchenko, {\it Usp. Fiz. Nauk} {\bf 164},
643 (1994).
\bibitem{Galynskii} M.V. Galynskii {\it et al.,} Zh. Eksp. Teor. Fiz. {\bf 95}, 1921 (1989);\\
M.V. Galynskii, S.M. Sikach, {\it Fiz. Elem. Chastits At. Yadra} {\bf 29}, 1133 (1998); \\
hep-ph/9910284.
\bibitem{Rogalyov} R.N. Rogalyov, {\it Teor. Mat. Fiz.} {\bf 101}, 384 (1994);
 {\it Int. J. Mod. Phys.} {\bf A11}, 3711 (1996).
\bibitem{Ballestrero} A. Ballestrero, E. Maina {\it Phys. Lett.} {\bf B350}, 225 (1995);
hep-ph/9911235;\\
E. Accomonado, A. Ballestrero, G. Passarino, {\it Nucl. Phys.} {\bf B476}, 3 (1996).
\bibitem{GPS} S. Jadah {\it et al., The Eur. Phys. Journal}, {\bf 22}, 423 (2001). 
\bibitem{CORE}  V.I. Borodulin, R.N. Rogalyov, S.R. Slabospitsky, hep-ph/9507456. 
\bibitem{Zuber} C. Itzykson, J.-B. Zuber, {\it Quantum Field Theory}, 
McGraw-Hill, 1980.
\bibitem{REDUCE} A. Hearn, Reduce User's manual, Version 3.3, 
Santa Monica: RAND Publication, 1987.
\bibitem{Likhoded} R.N. Rogalyov, hep-ph/0105187.
\bibitem{Cartan} E. Cartan, {\it "Theory of spinors",} Paris: Hermann, 1966.
\end{thebibliography}
\end{document}